\newcount\mgnf\newcount\tipi\newcount\tipoformule\newcount\greco

\tipi=2          
\tipoformule=0   

\global\newcount\numsec\global\newcount\numfor
\global\newcount\numapp\global\newcount\numcap
\global\newcount\numfig\global\newcount\numpag
\global\newcount\numnf

\def\SIA #1,#2,#3 {\senondefinito{#1#2}%
\expandafter\xdef\csname #1#2\endcsname{#3}\else
\write16{???? ma #1,#2 e' gia' stato definito !!!!} \fi}

\def \FU(#1)#2{\SIA fu,#1,#2 }

\def\etichetta(#1){(\veroparagrafo.\veraformula)%
\SIA e,#1,(\veroparagrafo.\veraformula) %
\global\advance\numfor by 1%
\write15{\string\FU (#1){\equ(#1)}}%
\write16{ EQ #1 ==> \equ(#1)  }}
\def\etichettaa(#1){(A\veraappendice.\veraformula)
 \SIA e,#1,(A\veraappendice.\veraformula)
 \global\advance\numfor by 1
 \write15{\string\FU (#1){\equ(#1)}}
 \write16{ EQ #1 ==> \equ(#1) }}
\def\getichetta(#1){Fig. \verafigura
 \SIA g,#1,{\verafigura}
 \global\advance\numfig by 1
 \write15{\string\FU (#1){\graf(#1)}}
 \write16{ Fig. #1 ==> \graf(#1) }}
\def\retichetta(#1){\numpag=\pgn\SIA r,#1,{\verapagina}
 \write15{\string\FU (#1){\rif(#1)}}
 \write16{\rif(#1) ha simbolo  #1  }}
\def\etichettan(#1){(n\verocapitolo.\veranformula)
 \SIA e,#1,(n\verocapitolo.\veranformula)
 \global\advance\numnf by 1
\write16{\equ(#1) <= #1  }}

\newdimen\gwidth
\gdef\profonditastruttura{\dp\strutbox}
\def\senondefinito#1{\expandafter\ifx\csname#1\endcsname\relax}
\def\BOZZA{
\def\alato(##1){
 {\vtop to \profonditastruttura{\baselineskip
 \profonditastruttura\vss
 \rlap{\kern-\hsize\kern-1.2truecm{$\scriptstyle##1$}}}}}
\def\galato(##1){ \gwidth=\hsize \divide\gwidth by 2
 {\vtop to \profonditastruttura{\baselineskip
 \profonditastruttura\vss
 \rlap{\kern-\gwidth\kern-1.2truecm{$\scriptstyle##1$}}}}}
\def\verapagina{
{\romannumeral\number\numcap}.\number\numsec.\number\numpag}}

\def\alato(#1){}
\def\galato(#1){}
\def\veroparagrafo{\number\numsec}\def\veraformula{\number\numfor}
\def\veraappendice{\number\numapp}
\def\verapagina{\number\pageno}\def\veranformula{\number\numnf}
\def\verafigura{{\romannumeral\number\numcap}.\number\numfig}
\def\verocapitolo{\number\numcap}\def\veranformula{\number\numnf}
\def\Eqn(#1){\eqno{\etichettan(#1)\alato(#1)}}
\def\eqn(#1){\etichettan(#1)\alato(#1)}

\def\Eq(#1){\eqno{\etichetta(#1)\alato(#1)}}
\def\eq(#1){\etichetta(#1)\alato(#1)}
\def\Eqa(#1){\eqno{\etichettaa(#1)\alato(#1)}}
\def\eqa(#1){\etichettaa(#1)\alato(#1)}
\def\dgraf(#1){\getichetta(#1)\galato(#1)}
\def\drif(#1){\retichetta(#1)}

\def\eqv(#1){\senondefinito{fu#1}$\clubsuit$#1\else\csname fu#1\endcsname\fi}
\def\equ(#1){\senondefinito{e#1}\eqv(#1)\else\csname e#1\endcsname\fi}
\def\graf(#1){\senondefinito{g#1}\eqv(#1)\else\csname g#1\endcsname\fi}
\def\rif(#1){\senondefinito{r#1}\eqv(#1)\else\csname r#1\endcsname\fi}
\def\bib[#1]{[#1]\numpag=\pgn
\write13{\string[#1],\verapagina}}

\def\include#1{
\openin13=#1.aux \ifeof13 \relax \else
\input #1.aux \closein13 \fi}

\openin14=\jobname.aux \ifeof14 \relax \else
\input \jobname.aux \closein14 \fi
\openout15=\jobname.aux
\openout13=\jobname.bib


\ifnum\tipoformule=1\let\Eq=\eqno\def\eq{}\let\Eqa=\eqno\def\eqa{}
\def\equ{}\fi


{\count255=\time\divide\count255 by 60 \xdef\hourmin{\number\count255}
        \multiply\count255 by-60\advance\count255 by\time
   \xdef\hourmin{\hourmin:\ifnum\count255<10 0\fi\the\count255}}

\def\oramin{\hourmin }

\def\data{\number\day/\ifcase\month\or january \or february \or march \or
april \or may \or june \or july \or august \or september
\or october \or november \or december \fi/\number\year;\ \oramin}

\setbox200\hbox{$\scriptscriptstyle \data $}

\newcount\pgn \pgn=1
\def\foglio{\number\numsec:\number\pgn
\global\advance\pgn by 1}
\def\foglioa{A\number\numsec:\number\pgn
\global\advance\pgn by 1}

\footline={\hss\tenrm\folio\hss}

\def\TIPIO{
\font\setterm=amr7 
\def \settepunti{\def\rm{\fam0\setterm}
\textfont0=\setterm   
\normalbaselineskip=9pt\normalbaselines\rm
}\let\nota=\settepunti}

\def\TIPITOT{
\font\twelverm=cmr12
\font\twelvei=cmmi12
\font\twelvesy=cmsy10 scaled\magstep1
\font\twelveex=cmex10 scaled\magstep1
\font\twelveit=cmti12
\font\twelvett=cmtt12
\font\twelvebf=cmbx12
\font\twelvesl=cmsl12
\font\ninerm=cmr9
\font\ninesy=cmsy9
\font\eightrm=cmr8
\font\eighti=cmmi8
\font\eightsy=cmsy8
\font\eightbf=cmbx8
\font\eighttt=cmtt8
\font\eightsl=cmsl8
\font\eightit=cmti8
\font\sixrm=cmr6
\font\sixbf=cmbx6
\font\sixi=cmmi6
\font\sixsy=cmsy6
\font\twelvetruecmr=cmr10 scaled\magstep1
\font\twelvetruecmsy=cmsy10 scaled\magstep1
\font\tentruecmr=cmr10
\font\tentruecmsy=cmsy10
\font\eighttruecmr=cmr8
\font\eighttruecmsy=cmsy8
\font\seventruecmr=cmr7
\font\seventruecmsy=cmsy7
\font\sixtruecmr=cmr6
\font\sixtruecmsy=cmsy6
\font\fivetruecmr=cmr5
\font\fivetruecmsy=cmsy5
\textfont\truecmr=\tentruecmr
\scriptfont\truecmr=\seventruecmr
\scriptscriptfont\truecmr=\fivetruecmr
\textfont\truecmsy=\tentruecmsy
\scriptfont\truecmsy=\seventruecmsy
\scriptscriptfont\truecmr=\fivetruecmr
\scriptscriptfont\truecmsy=\fivetruecmsy
\def \eightpoint{\def\rm{\fam0\eightrm}
\textfont0=\eightrm \scriptfont0=\sixrm \scriptscriptfont0=\fiverm
\textfont1=\eighti \scriptfont1=\sixi   \scriptscriptfont1=\fivei
\textfont2=\eightsy \scriptfont2=\sixsy   \scriptscriptfont2=\fivesy
\textfont3=\tenex \scriptfont3=\tenex   \scriptscriptfont3=\tenex
\textfont\itfam=\eightit  \def\it{\fam\itfam\eightit}%
\textfont\slfam=\eightsl  \def\sl{\fam\slfam\eightsl}%
\textfont\ttfam=\eighttt  \def\tt{\fam\ttfam\eighttt}%
\textfont\bffam=\eightbf  \scriptfont\bffam=\sixbf
\scriptscriptfont\bffam=\fivebf  \def\bf{\fam\bffam\eightbf}%
\tt \ttglue=.5em plus.25em minus.15em
\setbox\strutbox=\hbox{\vrule height7pt depth2pt width0pt}%
\normalbaselineskip=9pt
\let\sc=\sixrm  \let\big=\eightbig  \normalbaselines\rm
\textfont\truecmr=\eighttruecmr
\scriptfont\truecmr=\sixtruecmr
\scriptscriptfont\truecmr=\fivetruecmr
\textfont\truecmsy=\eighttruecmsy
\scriptfont\truecmsy=\sixtruecmsy
}\let\nota=\eightpoint}

\newfam\msbfam   
\newfam\truecmr  
\newfam\truecmsy 
\newskip\ttglue
\ifnum\tipi=0\TIPIO \else\ifnum\tipi=1 \TIPI\else \TIPITOT\fi\fi

\global\newcount\numpunt

\magnification=\magstephalf
\baselineskip=16pt
\parskip=8pt

\def\b{\beta}

\def\e{\epsilon}

\def\g{\gamma}
\def\k{\kappa}

\def\s{\sigma}

\def\z{\zeta}

\def\D{\Delta}
\def\L{\Lambda}
\def\G{\Gamma}

\def\del #1{\frac{\partial^{#1}}{\partial\l^{#1}}}

\def\E{{I\kern-.25em{E}}}
\def\N{{I\kern-.22em{N}}}
\def\M{{I\kern-.22em{M}}}
\def\R{{I\kern-.22em{R}}}
\def\Z{{Z\kern-.5em{Z}}}
\def\1{{1\kern-.25em\hbox{\rm I}}}
\def\eu{{1\kern-.25em\hbox{\sm I}}}

\def\C{{C\kern-.75em{C}}}
\def\P{{I\kern-.25em{P}}}

\def\del{\partial}


\def\AA{{\cal A}}

\def\FF{{\cal F}}

\def\SS{{\cal S}}

\def\MM{{\cal M}}

\def\LL{{\cal L}}

\def\chap #1#2{\line{\ch #1\hfill}\numsec=#2\numfor=1}

\def\un #1{\underline{#1}}

\def\ba{{\backslash}}

\def\wt{\widetilde}
\def\wh{\widehat}

\def\inn{\hbox{int}\,}


\newcount\foot
\foot=1
\def\note#1{\footnote{${}^{\number\foot}$}{\ftn #1}\advance\foot by 1}

\def\frac#1#2{{#1\over #2}}
\def\sfrac#1#2{{\textstyle{#1\over #2}}}
\def\text#1{\quad{\hbox{#1}}\quad}
\def\newpage{\vfill\eject}
\def\proposition #1{\noindent{\thbf Proposition #1:}}

\def\theo #1{\noindent{\thbf Theorem #1: }}
\def\lemma #1{\noindent{\thbf Lemma #1: }}
\def\definition #1{\noindent{\thbf Definition #1: }}

\def\proof{{\noindent\pr Proof: }}

\def\endproof{$\diamondsuit$}
\def\remark{\noindent{\bf Remark: }}
\def\thanks{\noindent{\bf Aknowledgements: }}
\font\pr=cmbxsl10
\font\thbf=cmbxsl10 scaled\magstephalf

\font\ch=cmbx12
\font\ftn=cmr8

\font\it=cmti10
\font\bf=cmbx10
\font\sm=cmr7
\font\vsm=cmr6

\font\tit=cmbx12
\font\aut=cmbx12
\font\aff=cmsl12
\def\s{\char'31}
\nopagenumbers
{$  $}
\vskip1.5truecm
\centerline{\tit THE LOW-TEMPERATURE PHASE OF KAC-ISING 
MODELS\footnote{${}^\#$}{\ftn Work
partially supported by:  Commission of the European Union
under contracts  CHRX-CT93-0411 and CIPA-CT92-4016,
 Czech Republic grants  \v{c} . 202/96/0731
and  \v{c}. 96/272.
}}
\vskip1.5truecm
\centerline{\aut Anton Bovier
\footnote{${}^1$}{\ftn e-mail:
bovier@wias-berlin.de}}
\vskip.1truecm
\centerline{\aff Weierstra\s {}--Institut}
\centerline{\aff f\"ur Angewandte Analysis und Stochastik}
\centerline{\aff Mohrenstra\s e 39}
\centerline{\aff  D-10117 Berlin, Germany}
\vskip.5truecm
\centerline{\aut  Milo\v{s} Zahradn\'\i k \footnote{${}^2$}{\ftn
e-mail: mzahrad@karlin.mff.cuni.cz}}
\vskip.1truecm
\centerline{\aff Department of Mathematics}
\centerline{\aff Charles University}
\centerline{\aff Sokolovsk\'a 83}
\centerline{\aff 186 00 Praha 8, Czech Republic }
\vskip1.5truecm\rm
\def\s{\sigma}
\noindent {\bf Abstract:} We analyse the low temperature phase of 
ferromagnetic Kac-Ising models in dimensions $d\geq 2$. We show that
if the range of interactions is $\g^{-1}$, then two disjoint 
translation invariant 
Gibbs states exist, if the inverse temperature $\b$ satisfies
$\b -1\geq \g^\k$, where $\k=\frac {d(1-\e)}{(2d+1)(d+1)}$, for any
$\e>0$. The prove involves the blocking procedure usual
for Kac models and also a  contour 
representation for the resulting long-range (almost) continuous
spin system which is suitable for the use of a variant of the 
Peierls argument. 


\newpage

\chap{1. Introduction}1

In 1963 Kac et al. [KUH] introduced a statistical mechanics model of particles
 interacting via long, but finite range interactions, i.e. through 
potentials 
of the form $J_\g(r)\equiv \g^d J(\g r)$, there $J$ is some function of
bounded support or rapid decrease (the original example was 
$J(r)= e^{-r}$) and $\g$ is a small parameter. These models were introduced 
as microscopic models for the van der Waals theory of the liquid-gas
transition. In fact, in the context of these models it proved 
possible to derive in a mathematically rigorous way the  
van der Waals theory including the Maxwell construction in the limit 
$\g\downarrow 0$. In mathematical terms, this is stated as the 
Lebowitz-Penrose theorem[LP]: The distribution of the density satisfies 
in the infinite volume limit
a large deviation principle with a rate function that, in the limit 
as $\g$ tends to zero, converges to the convex hull of the 
van der Waals free energy.  For a review of these results, see e.g. the 
textbook by Thompson [T].

Only rather recently there has been a more intense interest in the study of 
Kac models that went beyond the study of the global thermodynamic
potentials in the Lebowitz-Penrose limit, but that also considers 
the distribution of {\it local mesoscopic} observables. This program
has been carried out very nicely in the case of the Kac-Ising model in 
one spatial dimension 
by Cassandro, Orlandi, and Presutti [COP]. 
A closely related analysis had been performed earlier by Bolthausen and 
Schmock [BS]. These analysis can be seen as a rigorous derivation of 
a Ginzburg-Landau type field theory for these models. Very recently,
such an analysis was also carried out in a disordered version of the 
Kac Ising model, the  so-called Kac-Hopfield model by Bovier, Gayrard, and 
Picco [BGP1,BGP2]. 

An extension of this work to higher dimensional situations would of course
be greatly desirable. This turns out to be 
not trivial and,  surprisingly, even very elementary questions about
the Kac model in $d\geq 2$ are unsolved. One of them is the natural conjecture
that the critical inverse temperature $\b_{c}(\g)$ in the Kac model 
should converge, as $\g\downarrow 0$, to the mean-field critical 
temperature. This conjecture can be found e.g. in  a recent paper
by Cassandro, Marra, and Presutti [CMP]. In that paper 
a lower bound  $\b_{c}(\g)\geq 1+ b\g^2|\ln \g|$ is proven. A corresponding 
upper bound is only known in a very particular case where reflection 
positivity can be used [BFS]. 

In addressing this question one soon finds the reason for this 
unfortunate state of affairs. All the powerful modern methods for 
analyzing the low-temperature phases of statistical mechanics models,
like low-temperature expansions and the Pirogov-Sinai theory, 
have been devised in view of models with  short range 
(often nearest neighbor)
interactions, with possible longer range parts treated  as some 
nuisance that can be shown to be quite irrelevant. To deal with
the genuinely long-range interaction in Kac models, that is to
exploit their long range nature, these methods require substantial 
adaptation. The purpose of the present paper is to
help to develop adequate 
techniques to deal with this problem -- that beyond proving 
the conjecture of [CMP] will, hopefully, also provide a basis
for the analysis of disordered Kac models. (Together with possible other
means not touched by the presented article : most notably with
suitably developed expansion techniques for long range models).

The model we consider is defined as follows. We consider a measure space
$(\SS,\FF)$ where $\SS\equiv \{-1,1\}^{\Z^d}$ is equipped with the product
topology of the discrete topology on $\{-1,1\}$ and $\FF$ is the corresponding 
finitely generated sigma-algebra. We denote an element of $\SS$ by $\s$
and call it a spin-configuration. If $\L\subset\Z^d$, we denote by 
$\s_\L$ the restriction of $\s$ to $\L$. For any finite volume $\L$
we define the energy of the configuration $\s_\L$ (given the external
configuration $\s_{\L^c}$) as
$$
H_{\g,\L}(\s_\L,\s_{\L^c})\equiv - \sfrac 12\sum_{i,j\in \L,j\in \Z^d}
J_\g(i-j)\s_i\s_j-\sum_{i\in \L,j\not\in\L}J_\g(i-j)\s_i\s_j
\Eq(1.1)
$$
where 
$J_\g(i)\equiv \g^d J(\g i)$ and $J:\R^d\rightarrow \R$ is a function 
that satisfies $\int_{\R^d} dx J(x)=1$.    
For simplicity we will assume that $J$ has bounded support, but the 
extension  of our proof to more moderate assumptions on the decay 
properties 
of $J$ is apparently not too difficult. To be completely specific we will even 
choose $J(r)\equiv c_d\1_{|x|\leq 1}$ where $c_d$ normalizes the integral of
$J$ to one\note{The generic name $c_d$ will be used in the sequel for various
finite, positive constants that only depend on dimension.}. 
Here  $|\cdot|$ is most conveniently chosen as the sup-norm on $\R^d$.

Finite volume Gibbs measures (``local specifications'') are defined as usual as
$$
\mu_{\g,\b,\L}^{\eta}(\s_\L)\equiv
\frac 1{Z_{\g,\b,\L}^{\eta}} e^{-\b H_{\g,\L}(\s_\L,\eta_{\L^c})}
\Eq(1.2)
$$
where $Z_{\g,\b,\L}^{\eta}$ is the usual partition function.
Note that under our assumptions on $J$ the local specifications
for given $\L$
depend only on finitely many coordinates of $\eta$. 
An infinite volume Gibbs state  $\mu_{\g,\b}$ is a probability measure on 
$(\SS,\FF)$ that satisfies the DLR-equations
$$
\mu_{\g,\b} \mu_{\g,\b,\L}^{\cdot} = \mu_{\g,\b}
\Eq(1.3)
$$
Our first result will be the following

\theo{1} {\it Let $d\geq 2$. Then there exists a 
function $f(\g)$ with $\lim_{\g\downarrow 0}
f(\g)=0$ such that for all $\b> 1+f(\g)$, there exist at least two 
disjoint extremal infinite volume Gibbs states with local specifications
given by \eqv(1.2). Moreover, for $\g$ small enough, 
$f(\g)\leq \g^{\frac {1-\e}{(2d+2)(1+1/d)}}$ for arbitrary $\e>0$ }

\remark This theorem shows that the conjecture of [CMP] is correct. 
Together with Theorem 1 of [CMP] it implies that $\lim_{\g\downarrow 0}
\b_c(\g)=1$ in the Kac model. While completing this work we have received a
paper by M. Cassandro and E. Presutti [CP] in which the conjecture of 
[CMP] is also proven, but no explicit
estimate on the asymptotics of the function $f(\g)$ is given.
Their proof is rather different from ours. 
Although at the moment we  make use of the spin flip symmetry of
the model, the contour language we introduce is also intended
as a preparatory step for future use of the Pirogov-Sinai theory
for non-symmetric long range models.

We will in fact get more precise information on the infinite volume 
Gibbs measures in the course of the proof. 
This will be expressed in terms of the distribution 
of ``local magnetization'', $m_x(\s)$, defined on some suitable
length scale $1\ll\ell\ll\g^{-1}$. Given such scale $\ell$, we will partition 
the lattice $\Z^d$ into blocks, denoted by $x$ of side length $\ell$. We set
Identifying the block $x$ with its label $x\in \Z$, we could thus set
$$
x\equiv  \{i\in \Z^d \mid\, |i-\ell x|\leq \ell/2\}
\Eq(1.4)
$$
We then define
$$
m_x(\s)\equiv \frac 1\ell \sum_{i\in x}\s_i
\Eq(1.5)
$$
In the sequel we will assume that all finite volumes we consider are compatible
with these blocks, that is are decomposable into them. 
We will also assume that $\g\ell$ is an integer.   
For any volume $\L$ compatible with the block structure, we denote by $\MM_\L
\subset\FF_\L$
the sigma-algebra generated by the family of variables $\{m_x(\s)\}_{x\in \L}$.
The block variables will be instrumental in the proof of Theorem 1. However, 
they are also the natural variables to characterize the nature of typical 
configurations w.r.t. the Gibbs measure. We should note that this first step 
of passing to the variables $m_x(\s)$ is also used in [CP], in fact it is 
used in virtually all work on the Kac model. 
%
%

The remainder of this article is organized as follows. 
In Section 2 the distribution of the block spins are formally introduced
and the block-spin approximation of the Hamiltonian is discussed. 
In Section 3 we introduce our  notion of Peierls contours and prove our
theorem through variant of the Peierls argument [P].

\thanks 
We thank Errico Presutti and Marcio Cassandro for sending us 
a copy of their paper [CP] prior to publication. M. Zahradn\'\i k
also acknowledges useful discussions with E. Presutti on Kac models
in general and about their recent preprint in particular. 
We would like to thank also the home institutions of the authors 
and the Erwin Schr\"odinger Institute  in Vienna 
for  hospitality that made this
collaboration
possible.   

\newpage

\chap{2. Block spin approximation}2

All the questions we want to answer in our model will after all
concern the 
probabilities of events that are elements of the sigma-algebras
$\MM_V$ for finite volumes $V$. If $\AA\in\MM_V$ is such an event and
$\L\supset V$, we have the following useful identity
$$
\eqalign{
\mu^{\eta}_{\g,\b,\L}(\AA)&=\sum_{\s_{\L\ba V}}
\mu^{\eta}_{\g,\b,\L}(\s_{\L\ba V})\mu^{\s_{\L\ba V},\eta_{\L^c}}_{\g,\b,V}
(\AA)\cr
&=\sum_{\s_{\L\ba V}}
\mu^{\eta}_{\g,\b,\L}(\s_{\L\ba V})\sum_{{m_x,x\in V}\atop{ \{m_x\}\subset\AA}}
\mu^{\s_{\L\ba V},\eta_{\L^c}}_{\g,\b,V}\left(\{m_x\}_{x\in V}\right)
}
\Eq(2.1)
$$
The sum over $m_x$ runs of course over the values $\{-1,-1+2\ell^{-d},\dots,
1-2\ell^{-d},1\}$
Note that we may,  if $J$ has compact support,
assume without loss of generality that $\L$ is sufficiently large so that
the local specification $\mu^{\s_{\L\ba V},\eta_{\L^c}}_{\g,\b,V}$
does not depend on $\eta$. We will therefore drop the $\eta$ in 
this expression.

The main point which  makes the Kac-model special, is that the Hamiltonian 
is ``close'' to a function of the block spins. Namely, we may write
$$
\eqalign{
H_{\g,V}(\s_V,\s_{V^c})&= -\sfrac 12\sum_{x,y\in V} \sum_{i\in x,j\in y}
J_\g(i,j)\s_i\s_j-
\sum_{x\in V,y\in V^c} \sum_{i\in x,j\in y}
J_\g(i,j)\s_i\s_j\cr
&= -\sfrac 12\sum_{x,y\in V}J_\g(\ell(x-y)) \sum_{i\in x,j\in y}\s_i\s_j\cr
&
-\sum_{x\in V,y\in V^c}J_\g(\ell(x-y)) \sum_{i\in x,j\in y}\s_i\s_j\cr
& -\sfrac 12\sum_{x,y\in V} \sum_{i\in x,j\in y}
\left[J_\g(i-j)-J_\g(\ell(x-y))\right]\s_i\s_j\cr
&=H^{(0)}_{\g,\ell,V}(m_V(\s_V),m_{V^c}(\s_{V^c}))+
\D H_{\g,\ell,V}(\s_V,\s_{V^c})
}
\Eq(2.2)
$$
where we have set (recall that $J_{\g}(\ell x)=\ell^{-d}
J_{\ell\g}(x)$)
$$
H^{(0)}_{\g,\ell,V}(m_V,m_{V^c})\equiv 
-\ell^d\sfrac 12\sum_{x,y\in V}J_{\g\ell}(x-y) m_x m_y
-\ell^d\sum_{x\in V,y\in V^c}J_{\g\ell}(x-y) m_x m_y
\Eq(2.3)
$$
and
$$
\eqalign{
\D H_{\g,\ell,V}(\s_V,\s_{V^c})&= 
-\sfrac 12\sum_{x,y\in V} \sum_{i\in x,j\in y}
\left[J_\g(i-j)-J_\g(\ell(x-y))\right]\s_i\s_j\cr
&-\sum_{x\in V,y\in V^c} \sum_{i\in x,j\in y}
\left[J_\g(i-j)-J_\g(\ell(x-y))\right]\s_i\s_j
}
\Eq(2.4)
$$

\lemma {2.1} {\it For any $V\subset \Z^d$,
$$
\sup_{\s}\left| \D H_{\g,\ell,V}(\s_V,\s_{V^c})\right|\leq c_d \g\ell |V|
\Eq(2.5)
$$
where $c_d$ is some numerical constant that depends only on the dimension
$d$. }

\proof This fact is well-known and simple for all Kac models. In our case it 
follows from the observation that 
$
\left[J_\g(i-j)-J_\g(\ell(x-y))\right]=0$, unless  
$|x-y|\approx 1/ (\g\ell)$.\endproof

As consequence of Lemma 2.1 we get the following useful upper and lower bounds
for the distribution of the block spins:
$$
\eqalign{
\mu^{\s_{\L\ba V}}_{\g,\b,V}\left(m_V\right)&{{<}\atop{>}}
\frac{e^{-\b\ell^d H^{(0)}_{\g,\ell,V}(m_V,m_{V^c})   }
\prod_{x\in V}\E_\s \1_{m_x(\s)=m_x}}
{\sum_{m_V}{e^{-\b\ell^d   H^{(0)}_{\g,\ell,V}(m_V,m_{V^c})        }
\prod_{x\in V}\E_\s \1_{m_x(\s)=m_x}}}
e^{\pm \b c_d\g\ell |V|}
}
\Eq(2.6)
$$
Of course 
$$
\E_\s \1_{m_x(\s)=m_x}\cases{2^{-{\ell^d}}{{\ell^d}
\choose{\frac {1+m_x}2\ell^d}},&
if $\ell^d/ m_x/2\in \Z$\cr
0,&else}
\Eq(2.7)
$$
and thus, by Sterling's formula,
$$
  2^{-\ell^d}     {{\ell^d}\choose{\frac {1+m_x\ell^d}2\ell^d}}
  =e^{-\ell^dI(m_x)+O(\ln \ell)}
\Eq(2.8)
$$
where $I(m)$, for $m\in [-1,1]$ is 
$$
I(m)=\frac {1+m}2\ln(1+m)+\frac {1-m}2\ln (1-m)
\Eq(2.9)
$$                                          
Therefore  we define
$$
E_{\g,\b,\ell,V}(m_V,m_{V^c})\equiv -\sfrac 12 \sum_{x,y\in V}J_{\g\ell}(x-y)
m_x m_y -\sum_{x\in V,y\in V^c}J_{\g\ell}(x-y)
m_x m_y+
\b^{-1}\sum_{x\in V} I(m_x)
\Eq(2.10)
$$
to get 

\lemma{2.2}{\it For any finite volume $V$ and any configuration 
$m_V$, we have 
$$
 \mu^{\s_{\L\ba V}}_{\g,\b,V}\left(m_V\right){{<}\atop{>}}
\frac{e^{-\b\ell^d E_{\g,\b,\ell,V}(m_V,m_{V^c}(\s_{V^c}))}}
{\sum_{m_V}e^{-\b\ell^d E_{\g,\b,\ell,V}(m_V,m_{V^c}(\s_{V^c}))}}
e^{\pm\b c_d\g\ell |V|}
\Eq(2.11)
$$
}

\remark   $\ell$ will be chosen as tending to 
infinity as $\g$ tends to
 zero. The idea is that that  $ E_{\g,\b,\ell,V}$ is in a sense 
a ``rate function''; that is to say, 
$ E_{\g,\b,\ell,V}$ alone determines the measure since the 
residual entropy is only of the order $\frac {d\ln \ell}{\ell^d}|V|$.
The problem is that this is only meaningful when we consider 
events $\AA$ for which the minimal $ E_{\g,\b,\ell,V}$ is of order
$|V|$ above the ground state to make sure that neither
the residual entropy nor the error terms in \eqv(2.11) may invalidate the 
result.
We will have to work in the next section to define such events.

It is instructive to rewrite the functional $E_{\g,\b,\ell,V}$
in a slightly different form using that $-m_xm_y=\frac 12 (m_x-m_y)^2
-\frac 12(m_x^2+m_y^2)$ (we drop the 
indices ${\g,\b,\ell}$ henceforth but keep this dependence in mind).
We set 
$$
\eqalign{
\wt E_V(m_V,m_{V^c})&\equiv\sfrac 14\sum_{x,y\in V}J_{\g\ell}(x-y)
\left(m_x-m_y\right)^2
 +\frac 12\sum_{x\in V,y\in V^c}J_{\g\ell}(x-y)\left(m_x-m_y\right)^2\cr
&+\sum_{x\in V} f_\b(m_x)
}
\Eq(2.12)
$$
where $f_\b$ is the well-known free energy function of the Curie-Weiss model,
$$
f_\b\equiv
\left[\b^{-1} I(m_x)-\sfrac 12 m_x^2\right]
\Eq(2.12bis)
$$
Then 
$$
E_V(m_V,m_{V^c})=\wt E_V(m_V,m_{V^c})-C_V(m_{V^c})
\Eq(2.12ter)
$$
where
$$
C_V(m_{V^c})\equiv\frac 12\sum_{x\in V,y\in V^c}J_{\g\ell}(x-y)m_y^2
\Eq(2.12quater)
$$ 
depends only on the variables on $V^c$.

The form $\wt E_V$ makes nicely evident the fact that the energy functional 
favours configurations that are constant and close to the 
minima of the Curie-Weiss function
 $f_\b(m)$.


\vskip1.5cm

\chap{3. Peierls contours}3

In this Section we define an appropriate notion of Peierls-contours 
in our model and use this to proof Theorem 1 by a version of the
Peierls argument\note{While the proof of [CP] is also based on a 
Peierls argument,
their definition of Peierls contours is completely different from ours.}.
The general spirit behind the definition of
Peierls contours can be loosely characterized as follows: We want to 
define a family of local events that have the property that 
 at least one of them has to occur, if the effect of boundary conditions 
does not propagate to the interior of the system. Then one must 
show that the probability that any of these events occurs is small.
We will define such events in terms of the {\it block spin variables}
 $m_x(\s)$.
More precisely, since it is crucial for us to exploit that the new interaction 
is still long range\note{For that reason it is not possible to 
directly use the methods developed in [DZ] for studying low temperature
phases of
short range continuous spin models, although some of the ideas in that 
paper are used in our proof.},
contours will be defined in terms of the {\it local 
 averages}, $\phi_x(m)$, and the {\it local variances}, 
$\psi_x(m)$, defined through
$$
\phi_x(m)\equiv \sum_{y}J_{\g\ell} (x-y)m_y
\Eq(3.0)
$$
$$
\psi_x(m)\equiv \sum_{y}J_{\g\ell} (x-y)\left(m_y-\phi_y(m)\right)^2
\Eq(3.1)
$$
Then define the sets
$$
\un{ \wt \G}\equiv \left\{x \mid\left|\,|\phi_x(m)|-m^*(\b)\right|>{\z m^*(\b)}
\, \text {or} \psi_x(m) >(\z m^*(\b))^2\right\}
\Eq(3.2)
$$
where $m^*(\b)$ is the largest  
solution of the equation $x=\tanh\b x$, 
that is the location of the non-negative minimum of the function $f_\b$.
We recall (see e.g [E]) that $m^*(\b)=0$ if $\b\leq 1$, $m^*(\b)>0$ if 
$\b>1$, $\lim_{\b\uparrow \infty}m^*(\b)=1$ and 
$\lim_{\b\downarrow 1} \frac {(m^*(\b))^2}{3(\b-1)}=1$. 
To simplify notation we will 
write $m^*\equiv m^*(\b)$ in the sequel.
$\zeta,\hat z<1 $ will be chosen in a suitable way later.
Note that if the boundary conditions are such that 
say $\phi_x(m(\eta))\approx +m^*$, then, 
if the configuration near the origin is 
such that  $\phi_0(m(\s)<0$, there must be a region enclosing the origin on 
which $\phi$ takes the value zero and thus belongs to $\un{ \wt \G}$.
For a reason that will become clear later, in a first step we will 
regularize this set. For this we introduce a second blocking of the lattice, 
this time on the {\it scale of the range of the interaction}. 
The points $u$ of this 
lattice are identified with the blocks
$$
u\equiv\left \{x\in \Z^d|\,|x-u/(\g\ell)|\leq 1/(2\g\ell)\right\}
\Eq(3.2bis)
$$
just as in \eqv(1.4). We write in a natural way $u(x)$ for the label of the 
unique block that contains $x$. We will call sets that are unions of such 
blocks $u$ {\it regular sets}. 
We put
$$
\un{\G}_0\equiv \left\{ x\,| u(x)\cap \un{ \wt \G}\neq \emptyset\right\}
\Eq(3.2ter)
$$
For some positive integer $n\geq 1$ to be chosen later, we now set 
$$
\un\G \equiv\left\{x\mid \hbox{dist}(x,\un{  \G}_0)\leq n(\g\ell)^{-1}\right\}
\Eq(3.3)
$$
where $dist$ is the metric induced by the sup-norm on $\R^d$.
$n$ will depend on $\b$ and diverge as $\b\downarrow 1$.
The precise value of $n$ will be specified later in \eqv(3.33). Notice that
this definition assures that the set $\un\G$ is a regular set in  the sense 
defined above..
Connected components of the set set $\un \G$ together with  the specification 
of the values of $m_x$, $x\in \un \G$ are called contours and are denoted by 
$\G$.
For such a contour, we introduce the notion of its {\it boundary } $\del\G$, 
in the following sense:
$$
\del\G\equiv \left\{x\in \un\G\,\mid\,\hbox{dist}(x,\un\G^c)\leq
n(\g\ell)^{-1}\right\}
\Eq(3.03)
$$
Note that by our definition of $\G$ we are assured that $\del\G\cap \un{\G}_0
=\emptyset$.    
 We denote by
$$
D^\pm\equiv \left\{x \,\mid \, |\phi_x(m)\mp m^*|\leq {\z m^*}\right\}\cap 
\un\G^c
\Eq(3.4)
$$ 
and call these regions $\pm$-correct.
Each connected component of the boundary of $\G$ connects either to $D^+$ or
$D^-$. We will denote such connected components by $\del\G^+_i$ and 
$\del\G^-_i$, respectively.
    
For a connected set $\un \G $ we denote by $\inn\un\G$ the simply connected 
set obtained by ``filling up the holes'' of  $\un \G $. 
This set is called the interior  
of a contour. The boundary of  $\inn\un\G$ will be referred to as the 
exterior boundary of  $\un \G$. The connected component of $\del\G$ that
is also the  boundary of  $\inn\un\G$ will be called exterior boundary of 
$\G$ and denoted by $\del\G^{ext}$. 

The strategy to prove Theorem 1 is the usual one. First we observe that if 
boundary conditions are strongly plus, then in order to have that, say,
$|\phi_0(m)-m^*|>{\z m^*}$, it must be true that there exists a contour $\G$
such that $0\in\inn{\un\G}$ . Thus it suffices to prove that the probability 
of contours is sufficiently small. This will require a {\it lower bound} 
on the energy of {\it any configuration compatible with the existence of 
$\un\G$}, and an {\it upper bound} on a carefully chosen {\it reference 
configuration}
in which the contour is absent. We will show later (Lemma 3.8) 
that a lower bound on the 
energy can easily be given in terms of the functions $\phi$ and $\psi$, 
a fact that motivates the definition of $\un{\wt\G}$. The long range nature and
of the interaction and the fact that the $m_x$ are essentially continuous
variables require the construction of the extensive ``safety belts'' around
this set in order to assure an effective decoupling of the core of a 
contour from its exterior.
The crucial reason for the definition of contours through the nonlocal
functions $\phi$ and $\psi$ is however the fact that these are ``slowly 
varying'' functions of $x$ for any configuration $m$. Therefore, even if the
core $\un{\wt\G}$ is very "thin" (e.g. a single point), one can show
that on a much larger set $|\phi_0(m)-m^*|$ or $\psi_x(m)$ must 
still be quite large (e.g. half of what is asked for in
$\un{\wt\G}$). This guarantees that in spite of the very thick
``safety belts'' we must construct around  $\un{\wt\G}$, the energy 
of a contour compares nicely with its volume
$|\un{\G}|$.

We will now establish the ``decoupling'' properties.
For this we must establish some
properties of the configuration $m$ on $\del\G$ that minimizes $E_{\del\G}$
for given boundary conditions.

\definition  {3.1} {\it A configuration 
 $m^{opt}_V$ is called optimal if 
 $m^{opt}$ minimizes $E_{V}(m_{V},
m_{V^c})$ for a given configuration $ m_{V^c}$.
}

An important point is that away from $\wt{\un{\G}}$, due to our 
definition of contours 
configurations must be close to constant in the following sense:

\lemma {3.2} {\it  Assume that $\hbox{dist}(x,\un{\wt\G})>1/(\g\ell)$. Then 
\item{(i)}
$$
\sum_{y}J_{\g\ell}(x-y)\left(m_y\pm m^*\right)^2\leq 4\z^2(m^*)^2
\Eq(3.7bis)
$$
and 
\item {(ii)}for any $V\subset \un{\wt\G}$ 
$$
\sum_{y\in V}J_{\g\ell}(x-y) |m_y\pm m^*|\leq   2 \z m^* 
\sqrt{\sum_{y\in V}J_{\g\ell}(x-y)} 
\Eq(3.7ter)
$$
where the sign depends on whether $\phi_x(m)$ is positive or negative in 
the region.
}

\proof The proof of \eqv(3.7bis) is straightforward from the definition of 
$\un{\wt\G}$ in \eqv(3.2) and \eqv(3.7ter) follows from
\eqv(3.7bis) by the Schwartz inequality. \endproof

We will now establish properties of an optimal configuration on regular
 sets  with boundary conditions that satisfy properties \eqv(3.7bis)
and \eqv(3.7ter). 

\lemma {3.3} {\it Let $V$ be a regular set. 
Then there exists 
$\z_d>0$ depending only on the dimension $d$ such that if $m_{ V^c}$
 is a 
boundary conditions of $+$ type for which \eqv(3.7bis)
and \eqv(3.7ter) hold with $\z\leq \z_d$, then 
for all $x\in V$, $|m^{opt}_x-m^*|\leq m^*/2$. The corresponding 
statement holds for $-$ type boundary conditions.  
}

\proof  We see from \eqv(2.10) 
that we must have\note{We ignore the fact that $m_x$ takes only
discrete
values and look for the optimal solution in the space of real-valued
$m$. The point is that given such a solution, a discrete valued
approximation
can be constructed that differs in energy  by less than
 $|\un{\G}|/\ell^d$ which is negligibly small.}  for $y\in V$
$$
0=\frac {d}{dm_y}  
E_{V}(m_{V},
m_{V^c})= \b^{-1}I'(m_y)- \phi_y(m)
\Eq(3.8)
$$
\eqv(3.8) can be written as 
$$
m_y=\tanh\left(\b\phi_y(m)\right)
\Eq(3.9) 
$$
We may tacitly assume that $\phi_y(m)$ is positive 
(this assumption will be shown to be consistent).
Since $m^*$ is a stable fixpoint of the function $\tanh \b m$ that attracts 
all points on the positive half line, it follows that 
$
|\tanh\left(\b\phi_y(m)\right)-m^*|\leq |\phi_y(m)-m^*|
$
and in particular, if $ \phi_y(m)<m^*$, $\tanh\left(\b\phi_y(m)\right)>
\phi_y(m)$, while for $ \phi_y(m)>m^*$,
 $\tanh\left(\b\phi_y(m)\right)<
\phi_y(m)$. We will first show that $m_x^{opt}\geq m^*/2$. 
 Let $x\in V$ denote a point
where 
$$
m_x= \inf_{y\in V}\left\{m_y \,| m_y\leq m^*\right\}
\Eq(3.9ter)
$$
If $m_x=m^*$, there is nothing to proof. But if $m_x<m^*$, then 
\eqv(3.9ter) can only be satisfied if 
$\hbox{dist}(x,\del V)< 1/(\g\ell)$. For such points we can write
$$
\eqalign{
m_x-m^*&\geq \sum_{y\in V} 
J_{\g\ell}(x-y)(m_y-m^*) +\sum_{y\in V^c} J_{\g\ell}(x-y)(m_y-m^*)
\cr
&\geq (m_x-m^*)  \sum_{y\in V} J_{\g\ell}(x-y) -2\z m^* 
\sqrt{\sum_{y\in V^c} J_{\g\ell}(x-y)}
}
\Eq(3.9quater)
$$
where the second line follows by \eqv(3.7ter). Hence
$$
m_x-m^*\geq -\frac {2\z m^*}{\sqrt{\sum_{y\in V^c} J_{\g\ell}(x-y)}}
\Eq(3.9cinq)
$$
On the other hand, \eqv(3.9quater) holds for any other point $y\in V$
as well, and inserting this into the first line of \eqv(3.9quater) we
get
$$
m_x-m^*\geq(m_x-m^*)\sum_{y\in V}\sum_{z\in V} J_{\g\ell}(x-y) 
 J_{\g\ell}(y-z)-4\z m^*
\Eq(3.9six)
$$
Clearly we have won if either
$$
1-\sum_{y\in V}\sum_{z\in V} J_{\g\ell}(x-y)   
 J_{\g\ell}(y-z)\geq 8\z
\Eq(3.9sept)
$$
or 
$$
\sqrt{\sum_{y\in V^c} J_{\g\ell}(x-y)}\geq 4\z
\Eq(3.9oct)
$$
Due to the fact that $V$ is composed of cubes of sidelength of the 
range of the interaction, this follows from simple considerations
if $\z$ is smaller than some dimension dependent constant. (Here is the 
reason for our definition of $\un{\G}_0$). 
In fact,
$$
1-\sum_{y\in V}\sum_{z\in V} J_{\g\ell}(x-y)   
 J_{\g\ell}(y-z)=\sum_{y\in V^c} J_{\g\ell}(x-y) +
\sum_{y\in V}\sum_{z\in V^c} J_{\g\ell}(x-y)   
 J_{\g\ell}(y-z)
\Eq(3.9nov)
$$
The point is that the second term on the right hand side  of   \eqv(3.9nov)
cannot be too small as long as $\hbox{dist}(x,V^c)\leq 1/(\g\ell)$,
for regular $V$ (if $V$ is {\it not} regular, this statement does not
hold, of course; just consider a thin long spike entering into $V$ and let
$x$ be near the tip of the spike!). In fact, the worst situation here occurs 
if $x$ is at a distance $r/(\g\ell)$ from a ``corner'' of $V^c$. One easily 
verifies that even in this case 
$$
\eqalign{
& \sum_{y\in V}\sum_{z\in V^c} J_{\g\ell}(x-y)   
 J_{\g\ell}(y-z)\geq 2^{-(d+1)}\int_0^1 ds\,(r+s)^{d-1}(1-s)^d 
\cr
&\geq 2^{-(d+1)}\int_0^1 ds\, s^{d-1}(1-s)^d =2^{-(d+2)}\frac
{((d-1)!)^2}{(2d-1)!}
}
\Eq(3.9dez)
$$
so that \eqv(3.9oct) is verified if $4\zeta$ is smaller than this number.
The numerical value of that bound can of course be improved, 
but we do not seek to do that.

Having established that  $m_x\geq m^*/2$ in $V$, a trivial computation 
shows that our starting assumption that $\phi_x(m)>0$ 
is also verified. 
Thus we have proven that $m_x^{opt}\geq m^*/2$. In the same way one shows also
that $m_x^{opt}\leq 3m^*/2$ which concludes the proof 
of the lemma.\endproof

In the sequel the notion of $n$-layer set defined in the following definition 
will be convenient.

\definition  {3.4} {\it A regular set  $V$ is called a $n$-layer annulus, 
if there
it is of the form
$$
V=\left\{ x\in \wt V^c\,\mid\,\hbox{dist}(x,\wt V)\leq n(\g\ell)^{-1}\right\}
\Eq(3.5)
$$
for some connected set $\wt V$ that is composed of blocks $u$. The sets
$$
V_k\equiv  \left\{ x\in \wt V^c\,\mid\ (k-1)(\g\ell)^{-1}<
\hbox{dist}(x, \wt V)\leq k(\g\ell)^{-1}\right\}
\Eq(3.6)
$$
are called the $k-th$ layers of $V$. 
}
 
Note that the sets $\del\G$ are by their definition $n$-layer sets. 

We are interested in some properties of optimal 
configurations on $n$-layer sets. For this we will 
use the following simple fact   about 
the function $f_\b$, that may be found e.g. in [BG]

\lemma {3.5} { \it Let $f_\b(m)=\b^{-1} I(m)-\frac 12 m^2$.  Then, for all 
$m\in [-1,1]$
 $$
f_\b(m)-f_\b(m^*)\geq c(\b)\left(|m|-m^*\right)^2
\Eq(3.10)
$$
where 
$$ 
c(\b)\equiv \frac {\ln\cosh(\b m^*)}{\b (m^*)^2}-\frac 12
\Eq(3.11)
$$
has the                       
property that 
$c(\b)>0$ for all $\b>1$ and
$$
\lim_{\b\downarrow 1} \frac {c(\b)}{ (m^*)^2}=\frac 1{12}
\Eq(3.12) 
$$
}
From  this we will derive the following Lemma (The analog of this Lemma 
for short range and 
purely quadratic Hamiltonians 
appeared already in [DZ]).

\lemma {3.6} {\it Let $V$ be an $n$-layer set with $n\geq r/c(\b)$. Then 
there exists a layer $V_k$ in $V$ such that 
$$
\sum_{x\in V_k}\left(m_x^{opt}\right)^2 \leq 2^{-r} \frac 18( m^*)^2 
(|V_1|+|V_n|)
\Eq(3.13)
$$
}

\proof Let us set $u_x\equiv |m_x|-m^*$ and  and  use the abbreviation
$$
\|u_{V_k}\|_2^2
\equiv \sum_{x\in V_k}\left(u_x\right)^2
\Eq(3.14)
$$
and analogously for other functions.
 Then it is obvious from \eqv(2.12) that for any configuration,
$$
\wt E_{V\ba V_1\ba V_2}(m_{V\ba V_1\ba V_n},m_{V_1\cup V_n})
\geq  
\sum_{k=2}^{n-1}
c(\b) \|u_{V_k}\|_2^2 
+\sum_{x\in V\ba V_1\ba V_n}f_\b(m^*)
\Eq(3.15)
$$

On the other hand, we may consider a configuration that equals
 $m^{opt}$ on $V_1$ and $V_n$ and has $m_x=m^*$ for all $x\in 
V\ba V_1\ba V_n$. 
For this configuration 
$$
\eqalign{
\wt E_{V\ba V_1\ba V_2}(m_{V\ba V_1\ba V_n}=m^*,m^{opt}_{V_1\cup V_n})&=
\frac 12\sum_{{x\in V\ba V_1\ba V_n}
\atop {y\in V_1\cup V_n}} J_{\g\ell}(x-y)\left(m^{opt}_y-m^*\right)^2  +
\sum_{x\in V\ba V_1\ba V_n}f_\b(m^*)\cr
}
\Eq(3.15bis)
$$
By the definition of $m^{opt}$, it must thus be true that
$$
\eqalign{
0&\geq \wt E_{V\ba V_1\ba V_2}(m^{opt}_{V\ba V_1\ba V_n},m^{opt}_{V_1\cup V_n})
-\wt E_{V\ba V_1\ba V_2}
(m^{opt}_{V\ba V_1\ba V_n}=m^*,m^{opt}_{V_1\cup V_n})\cr
&\geq 
\sum_{k=2}^{n-1}
c(\b) \|u_{V_k}\|_2^2 
- \frac 12\sum_{{x\in V\ba V_1\ba V_n}
\atop {y\in V_1\cup V_n}} J_{\g\ell}(x-y)  \left(m^{opt}_y-m^*\right)^2
\cr
&\geq\sum_{k=2}^{n-1}
c(\b) \|u_{V_k}\|_2^2-
\frac 12\left(\|u^{opt}_{V-1}\|_2^2+\|u^{opt}_{V_n}\|_2^2\right)
}
\Eq(3.16)
$$
Thus, for any $q<n/2$, we have 
$$
q c(\b) \inf_{k=2}^{q+1}  \left[\|u^{opt}_{V_k}\|_2^2+
 \|u^{opt}_{V_{n+1-k}}\|_2^2\right]\leq 
 \sum_{k=2}^{n-1}
c(\b) \left[\|u^{opt}_{V_k}\|_2^2+
 \|u^{opt}_{V_{n+1-k}}\|_2^2\right]
\leq 
\sfrac 12 \|u^{opt}_{V_1}\|_2^2 +\sfrac 12 \|u^{opt}_{V_1}\|^2_2
\Eq(3.18)
$$
from where 
$$
 \inf_{k=2}^{q+1}  \left[\|u_{V_k}\|_2^2+
 \|u_{V_{n+1-k}}\|_2^2\right]
\leq \frac {1}{2q c(\b)}  
\left[\|u^{opt}_{V_1}\|_2^2 +\|u^{opt}_{V_1}\|^2_2\right]
\Eq(3.19)
$$
If $q$ is chosen as the smallest integer greater than $1/c(\b)$
this shows that there exist $2\leq k\leq q+1$ such that 
$$
 \left[\|u_{V_k}\|_2^2+
 \|u_{V_{n+1-k}}\|_2^2\right]\leq \frac 12 
\left[\|u^{opt}_{V_1}\|_2^2 +\|u^{opt}_{V_1}\|^2_2\right]
\Eq(3.20)
$$
Iterating this construction, and using that 
by Lemma 3.3
$$
\frac 12 \|u^{opt}_{V_1}\|_2^2 +\sfrac 12 \|u^{opt}_{V_1}\|^2_2
\leq \frac 18( m^*)^2 \left(|V_1|+|V_n|\right)
\Eq(3.20bis)
$$
we arrive at the assertion of the lemma. \endproof

We are now ready to construct our reference configuration and give an upper 
bound on its energy.  
For given contour $\G$ and compatible external configuration $m$ on $\un{\G}^c$
we call $m^{opt}$ the configuration on $\un\G$ that minimizes the energy
for a given core $\un{\wt\G}$. Clearly such a configuration is
also an optimal configuration on $\del\G$ in the sense of Definition 3.1.
Thus by Lemma 3.2 we know that in each connected component $\del\G^\pm_i$ of the boundary of
$\G$ there exists a layer $\LL^\pm_i$ of thickness 
$1/(\g\ell)$ in $\del\G_i^\pm$ such that $\|m_{\LL}^{opt}\mp m^*\|_2^2
\leq 2^{-r}\sfrac 18 (m^*)^2\frac 12 [|V_1(\del\G_i^\pm)|+|V_n(\del\G_i^\pm)|]$
For given  $\LL^\pm_i$ we decompose $\del\G_i^\pm$ into the two sets
$$
\del\G_{i,in}^\pm \equiv \left\{ x\in \del\G_i^\pm\ba\LL_i^\pm
\,\mid, \hbox{dist}(x,D^\pm)>\hbox{dist}(\LL_i^\pm,D^\pm)\right\}
\Eq(3.21)
$$
and 
$$
\del\G_{i,out}^\pm \equiv \del\G_i^\pm\ba\del\G_{i,in}^\pm
\Eq(3.22)
$$

Without loss of generality we assume that the exterior boundary of our 
contour is attached to the $+$-correct region.  
We now define the reference configuration $m^{ref}$ 
$$
m^{ref}_x\equiv\cases{m_x^{opt},&if $x\in \del\L_{i,out}^+$ \cr
                       -m_x^{opt},&if $x\in \del\L_{i,out}^-$ \cr
                       m^*,& for all other $x\in\un\G$\cr
                       m_x,& for $x\in  D^+$\cr
                       -m_x<& for $x\in D^-$
}
\Eq(3.23)
$$

\lemma {3.7} {\it Let $m^{ref}$ be defined in \eqv(3.23). Then for any 
compatible external configuration we have that 
$$
\eqalign{
\wt E_{\un\G}\left(m^{ref}_{\un\G},m^{ref}_{\un\G^c}\right)
&\leq  \sum_{i,\pm}
\wt E_{\del\G_{i,out}^\pm}\left(m^{opt}_{\del\G_{i,out}^\pm},m_{\un\G^c}
\right)
+\sum_{i,\pm} 2^{-r}\sfrac 18 (m^*)^2 
[|V_1(\del\G_i^\pm)|+|V_n(\del\G_i^\pm)|]\cr
&
+\sum_{x\in \un\G\ba\del\G_{out}} f_\b(m^*)
}
\Eq(3.24)
$$
}

\proof The proof of this estimate is obvious from the definition of
$m^{ref}$ and Lemma 3.6. Note that in the terms 
$\wt E_{\del\G_{i,out}^\pm}\left(m_{\del\G_{i,out}^\pm},m_{\un\G^c}
\right)$ the interaction energy between $\del\G_{i,out}^\pm$ and $\del
\G_{i,in}^\pm$ is not counted. \endproof 

Of course the configuration $m^{ref}$ does not contain the contour $\G$. It 
remains to find a lower bound on the energy of any configuration
$m$ that does contain a contour with given $\un{\wt\G}$. 

To do this, we use the following observation.

\lemma {3.8} {\it Let $U,V,W\subset\Z^d$ be any three disjoint sets
such that for all $y\in U\cup W$, $\sum_{x\in U\cup W\cup V}J_{\g\ell}(x-y)=1$.
and for any $y\in U$ $\sum_{x\in U\cup W}J_{\g\ell}(x-y)=1$
Then 
$$
\eqalign{
\wt E_{ V\cup U\cup W}(m_{ V\cup U\cup W},m_{(V\cup U\cup W)^c})
&\geq \sfrac 14 \sum_{x\in U} \psi_x(m)+\sfrac 12\sum_{x\in U\cup W}\left[
f_\b(m_x)+f_\b\left(\phi_x(m)\right)\right]\cr
&+\sum_{x\in V}f_\b(m^*)
}
\Eq(3.25)
$$
}

\proof  The proof of this lemma is a simple, but, mainly because 
of boundary effects, somewhat lengthy 
computation that we do not wish to reproduce here. 
To get the idea, note that in infinite volume we have (formally)
$$
\eqalign{
&-\sfrac 12\sum_{x,y} m_xm_y J_{\g\ell}(x-y)+\b^{-1}\sum_{x}I(m_x)\cr
&=-\sfrac 12\sum_{x} m_x\phi_x(m)+\b^{-1}\sum_{x}I(m_x)\cr
&= \sum_{x}\left[\frac {(m_x-\phi_x(m))^2}4 -\frac {m_x^2}4-
\frac {(\phi_x(m))^2}4+\frac 12\b^{-1}I(m_x)+ \frac 12\b^{-1}\phi_x(I(m))
\right]
}
\Eq(3.25bis)
$$
where we have put $\phi_x(I(m))=\sum_{y}J_{\g\ell}(x-y) I(m_y)$. The last line 
is obtained by inserting the identity $1=\sum_{y}J_{\g\ell}(x-y)$ in the 
$I(m)$ term and changing the order of summation in the resulting double sum.
Using the same trick for the first term in the last line of \eqv(3.25bis),
and using that, since $I$ is a convex function, $\phi_x(I(m))\geq I(\phi_x(m)$,
one gets that
$$
\sum_{x}\left[\frac 14 \psi_x(m)+\frac 12 f_\b(\phi_x(m))+\frac 12 f_\b(m_x)
\right]
\Eq(3.25ter)
$$
is a lower bound for \eqv(3.25bis). 
Trying to repeat this computation in finite volume and carefully dealing with 
the boundary terms leads to the more complicated looking formula \eqv(3.25).
\endproof

The main point in the estimate \eqv(3.25) is that it allows to bound the energy
of a configuration from below in terms of $\phi_x(m)$ and $\psi_x(m)$ alone. 
Namely, taking for $V$ and $U\cup W$ the layers $\LL^\pm_i$ and the regions 
``within''  $\LL^\pm_i$, we see that for any configuration  
$$
\eqalign{
\wt E_{\un\G}\left(m_{\un\G},m_{\un\G^c}\right)
&\geq 
\sum_{i,\pm}\wt E_{\del\G_{i,out}^\pm}\left(m_{\del\G_{i,out}^\pm},m_{\un\G^c}
\right)
\cr
&+\sfrac 12\sum_{x\in \un{\G}\ba \del\G_{out}}  \left[
f_\b\left(\phi_x(m)\right)- f_\b(m^*)\right]
+\sfrac 14 \sum_{{x\in \un\G\ba\del\G_{out}}\atop{\hbox{\vsm dist}
(x,\del\G_{out})>
1/(\g\ell)}}\psi_x(m)\cr
&+\sum_{x\in \un\G\ba\del\G_{out}} f_\b(m^*)
\cr
}
\Eq(3.28ter)
$$

Next we show that  bot $\phi_x(m)$ and $\psi_x(m)$ has nice continuity properties.

\lemma {3.9} {\it There exists a finite constant $c_d$ depending only 
on the dimension $d$ such that for any contour $\G$, if
$\un{\widehat\G}$ denotes the set
$$
\un{\widehat \G}\equiv \left\{y\,\mid\,\hbox{dist}(y,\un{\wt\G})\leq 
\frac {{(\z m^*)^2}}{8c_d \g\ell}
\right\}
\Eq(3.29)
$$
then for all $y\in \un{\widehat \G}$, $||\phi_y(m)|-m^*|\geq 
\frac {{\z m^*}}2$, or $\psi_x(m)\geq \frac {(\z m^*)^2}2$.
}

\proof Since $|m_x|\leq 1$, it is a simple geometric fact that
$$
|\phi_x(m)-\phi_y(m)|\leq c_d |x-y|\g\ell
\Eq(3.30)
$$
and 
$$
|\psi_x(m)-\psi_y(m)|\leq 4c_d|x-y|\g\ell
\Eq(3.30bis)
$$
for some geometry dependent constant $c_d$. Since on $\un{\wt\G}$,
$|\phi_x(m)| \geq {\z m^*}$ or $\psi_x(m)\geq (\z m^*)^2$, the assertion of the lemma follows. \endproof

\remark The estimates of Lemma 3.9 are very crude. We expect that they 
can be improved considerably. 
 
A further simple geometric consideration shows on the other hand
that $\un{\wh\G}$ cannot be too small compared to $\un\G$, namely

\lemma {3.10} {\it There exists a numerical constant 
$c_d'$ depending only on the dimension $d$ such that 
for any contour $\G$, we have that 
$$
|\un \G|\leq c_d'  
\frac {(n+1)^d}{ (\z m^*)^{2d}} |\un {\widehat\G}|
\Eq(3.31)
$$
}
\proof Note that $\frac {|\un \G|}{|\un {\widehat\G}|}$ 
is maximal if $\un {\widehat\G}$ consists of a single point 
in which case \eqv(3.31) is obvious.\endproof

Combining the upper bound on the energy of $m^{ref}$ from Lemma 3.7 
with the lower 
bound \eqv(3.28ter) obtained from Lemma 3.8 applied 
for the optimal configuration,
using the fact that that $E$ and $\wt E$ differ only by a
constant that depends only on boundary conditions, and finally employing 
Lemma 3.10
we arrive at 

\proposition {3.11} {\it Let $\G=(\un \G, m)$ be a contour with fixed 
$\un \G$. Then there
exists a reference configuration $m^{ref}$ in which $\G$ does not occur
such that     
$$
E_{\un\G}\left(m_{\un\G},m_{\un\G^c}\right)-
E_{\un\G}\left(m^{ref}_{\un\G},m^{ref}_{\un\G^c}\right)
\geq \frac 1{8} \frac {c(\b)}{c_d}\frac { (\z m^*)^{2d+2}}{(n+1)^d} |\un\G|
-\sfrac 18 (m^*)^22^{-n c(\b)}|\un\G|
\Eq(3.32)
$$
where $c_d$ is a finite dimension-dependent constant and $c(\b)$ 
is the constant from \eqv(3.11).}

\proof We bound $E_{\un\G}(m_{\un\G},m_{\un{\G}^c})$ from below by the the
 the corresponding energy of the configuration $m$ of lowest energy for 
given $\un \G$; on the belt of the contour this provides a optimal 
configuration in the sense of Definition 3.1. The same configuration
is used in the construction of $m^{ref}$. After the obvious cancelations
and using \eqv(3.31) and the fact that
$c(\b)\leq 1$, we get the assertion of the proposition. \endproof

We must now begin to choose our parameters. 
We want the Peierls condition, i.e. that the coefficient 
of $|\un \G|$ in \eqv(3.32) is positive and as large as possible.
The most convenient choice appears to choose $n$ in such a way that
$$ 
2^{-n c(\b)}=\frac 12 \frac {c(\b)(\z m^*)^{2d}}{c_d (n+1)^d}
\Eq(3.33)
$$
Calling the solution\note{By this we will of course understand the 
smallest integer larger than or equal to the ``real'' solution}
 of this equation $n^*$, 
we get the Peierls estimate
$$
E_{\un\G}\left(m_{\un\G},m_{\un\G^c}\right)-
E_{\un\G}\left(m^{ref}_{\un\G},m^{ref}_{\un\G^c}\right)
\geq 
\frac 1{16} \frac {c(\b)(\z m^*)^{2d+2}}{c_d (n+1)^d}
\Eq(3.34)
$$ 
It is not difficult to verify that 
$$
n^*\leq C\frac 1{c(\b)}\ln \left[\frac {c(\b)(\z m^*)^{2d}}{2 c_d}\right]
\Eq(3.34bis)
$$
for some numerical constant $C$, if $c(\b)$ is sufficiently small. 
 
This estimate on the energy difference will only be useful for us if
it is large compared to the error terms arising from the block approximation.
That is to say, we must make sure that 
$$
\frac 1{16} \frac {c(\b)(\z m^*)^{2d+2}}{c_d (n+1)^d}
> c_d\g\ell 
\Eq(3.35)
$$
(the two $c_d$ in this formula are a priori not the same quantities).
This gives a relation between  temperature dependent quantities on the one hand
and $\g \ell$ on the other. It does not impose any choice on the parameter 
$\ell$.  This arises from the last condition, the comparison between 
the energy of a contour and the entropy, i.e. the number of configurations
$m$ on $\un\G$ and of shapes $\un\G$ with fixed volume $|\un\G|$. 
Even the crudest estimate shows that this number is smaller than
$
\ell^{d |\un\G|} C^{d |\un\G|}
$
so that \eqv(3.35) is complemented by the 
condition 
$$
\b\ell^d\left[
\frac 1{16} \frac {c(\b)(\z m^*)^{2d+2}}{c_d (n+1)^d}
- c_d\g\ell\right] >d\ln \ell +\ln C  
\Eq(3.36)
$$
which requires $\ell$ to be sufficiently large. 
In fact we may choose $\ell$ as
$$
\ell=\g^{-1} \frac 1{c_d}\frac 1{32} 
\frac {c(\b)(\z m^*)^{2d+2}}{c_d (n+1)^d}
\Eq(3.37)
$$
which inserted into 
\eqv(3.36) gives the final condition of $\b$ in terms of $\g$.
It is clear that for any $\b>1$, i.e. $c(\b)>0$ and $m^*>0$, this 
condition can be verified by choosing $\g$ sufficiently small. 
Thus using Lemma 2.2 we proved the analog of the 
Peierl's argument here, namely that
the probability of a given contour $\G$ is smaller than
$\exp(-c|\un\G||\ln \ell|)$ which in turn implies that the probability
that the origin is in the interior of a contour is close to zero
(in fact of the order $\exp(-c\b n^d |\ln \ell|)$).
Moreover, 
by inserting the asymptotic behaviour of $m^*$ and $c(\b)$, one verifies 
easily that if we put 
$$
\b-1=\g^{\frac {1-\e}{(2d+2)(1+1/d)}}
\Eq(3.38)
$$
for arbitrary $\e>0$, then \eqv(3.36) is verified when $\g$ is sufficiently
small. This gives thus the claimed bound on the behaviour of the critical
temperature as $\g$ tends to $0$.

This concludes the proof of Theorem 1.\endproof\endproof

\remark Let us recall some consequences of what we have just proven: 
if $V$ denotes the union of the interiors of all the contours of a
given configuration than the Gibbs probability of the event 
$$
\hbox{dist}(i,V^c)\geq r 
\Eq(3.39)
$$
is independent of the choice of the point $i\in \Z^d$ and behaves like
$\exp(-C r)$ where $C=C(\b,\g)$. This implies for example the
following statement: The probability of the event that the support of 
all contours surrounding a given point is infinite is equal to zero.
One could even refine such a statement, giving a more precise meaning
to the intuitive idea that ``almost all configurations (of the
mesoscopic
observables $m$) in the
translation
invariant  $+$ Gibbs ensemble have their local averages (in the 
sense of the variables $\phi_x(m)$) in the vicinity of $m^*$ except of
some (rare, but uniformly distributed throughout the lattice)
``islands''. (This is the appropriate rephrasing  of the statement 
in Sinai's book [S]). 

\newpage

\chap{References}0

\item{[BGP1]} A. Bovier, V. Gayrard, and P. Picco,
``Large deviation principles for the Hopfield model and the
Kac-Hopfield model'', Prob. Theor. Rel. Fields {\bf 101}, 511-546 (1995).
\item {[BGP2]} A. Bovier, V. Gayrard, and P. Picco,
``Distribution of overlap profiles in the one dimensional Kac-Hopfield
model'', WIAS-preprint 221, submitted to Commun. Math. Phys. (1996).
\item{[BFS]} J. Bricmont, J.R. Fontaine, and E. Speer, ``Perturbation about
the mean field critical point'', Commun. Math. Phys. {\bf 86}, 337-362
(1982). 
\item{[BS]} E. Bolthausen and U. Schmock, ``Convergence of path measures 
arising from a mean field or polaron type interaction'',
  Prob. Theor. Rel. Fields {\bf 95}, 283-310 (1993).
\item{[CMP]} M. Cassandro, R. Marra, and E. Presutti, 
``Corrections to the critical temperature in 2d Ising systems with 
Kac potentials'', J. Stat. Phys. {\bf 78}, 1131-1138 (1995).
\item{[COP]} M. Cassandro, E. Orlandi, and E. Presutti,
``Interfaces and typical Gibbs configurations for one-dimensional
Kac potentials'', Prob. Theor. Rel. Fields {\bf 96}, 57-96 (1993).
\item{[CP]}  M. Cassandro and E. Presutti, ``Phase transitions in
  Ising systems with long but finite range'', preprint (1996).
\item{[DZ]} R.L. Dobrushin and M. Zahradn\'\i k, ``Phase diagrams for 
continuous-spin models: an extension of the Pirogov-Sinai theory'', in
``Mathematical Problems of Statistical Mechanics'', R.L. Dobrushin,
Ed. (D. Reidel Publishing Company, Dordrecht, 1986).
\item{[E]} R.S. Ellis, ``Entropy, large deviations, and statistical
mechanics'',  (Springer-Verlag, Berlin, 1985).
\item{[KUH]} M. Kac, G. Uhlenbeck, and P.C. Hemmer, ``On the 
van der Waals theory of liquid-vapour equilibrium. I. Discussion of a 
one-dimensional model'', J. Math. Phys. {\bf 4}, 216-228 (1963);''II.
Discussion of the distribution functions'', J. Math. Phys. {\bf 4},
229-247 (1963); ``III. Discussion of the critical region'', J. Math.
Phys. {\bf 5}, 60-74 (1964).
\item {[LP]} J. Lebowitz and O. Penrose, ``Rigorous treatment of the 
van der Waals Maxwell theory of the liquid-vapour transition'',
J. Math. Phys. {\bf 7}, 98-113 (1966).
\item{[P]} R. Peierls, ``On the Ising model of ferromagnetism'',
Proc. Cambridge Phil. Soc. {\bf 32}, 477-481 (1936).
\item{[S]} Ya.G. Sinai, ``Theory of phase transitions. Rigorous
  results'',
(Pergamon Press, New York, 1982)
\item{[T]} C.J. Thompson, ``Classical equilibrium statistical 
mechanics'', (Clarendon Press, Oxford, 1988).

\end